\documentclass[prl,twocolumn]{revtex4}
\usepackage{amssymb}
\usepackage{color}

\usepackage{amsthm}
\usepackage{graphicx}
\usepackage{dcolumn}
\usepackage{bm}
\usepackage{subfigure}
\usepackage{amssymb}
\usepackage{verbatim}
\usepackage{amscd}
\usepackage{amssymb}
\usepackage{amsmath, amsfonts}
\usepackage{setspace}
\usepackage[]{graphicx}        
\usepackage{amsthm}
\usepackage{enumerate}

\newcommand{\ba}[1]{\begin{align*} #1 \end{align*}}

\theoremstyle{plain}

\theoremstyle{plain}

\theoremstyle{plain}
 
\theoremstyle{plain}

\theoremstyle{remark}

\theoremstyle{conjecture}

\theoremstyle{observation}

\theoremstyle{definition}

\theoremstyle{corollary}

\theoremstyle{definition}

\theoremstyle{definition}

\theoremstyle{assumption}

\theoremstyle{definition}

\theoremstyle{problem}

\theoremstyle{fact}

\begin{document}

\title{Violation of the Arrhenius law below the transition temperature}
\author{Beni Yoshida}
\affiliation{Institute for Quantum Information and Matter, California Institute of Technology, Pasadena, California 91125, USA
}

\date{\today}
\begin{abstract}
When interacting spin systems possess non-zero magnetization or topological entanglement entropy below the transition temperature, they often serve as classical or quantum self-correcting memory. In particular, their memory time grows exponentially in the system size due to polynomially growing energy barrier, as in the 2D Ising model and 4D Toric code. Here, we argue that this is not always the case. We demonstrate that memory time of classical clock model (a generalization of ferromagnet to q-state spins) may be polynomially long even when the system possesses finite magnetization. This weak violation of the Arrhenius law occurs above the percolation temperature, but below the transition temperature, a regime where excitation droplets percolate the entire lattice, yet the system retains a finite magnetization. We present numerical evidences for polynomial scaling as well as analytical argument showing that energy barrier is effectively suppressed and is only logarithmically divergent. We also suggest an intriguing possibility of experimentally observing the precession of magnetization vectors at experimentally relevant time scale. 
\end{abstract}
\maketitle

\emph{Introduction --}
One of the challenges in building a large-scale quantum computing architecture is the fact that quantum entanglement decays easily, and one needs to protect a qubit from thermal decoherence. An important question in quantum information science concerns the feasibility of self-correcting quantum memory, a hypothetical memory device that would store a logical qubit for arbitrary long time without any error-correction~\cite{Bravyi09, Hamma09, Beni10, Alicki10, Beni10b, Beni11, Haah11, Michnicki12, Kim12, Landon-Cardinal13, Pedrocchi13, Beni13}. This problem is also closely related to another condensed matter physics question whether topological order may exist at finite temperature or not~\cite{Bravyi09, Beni11, Hastings11, Castelnovo08, Bravyi10b}.

The physical mechanism for self-correcting quantum memory and its relation to topological order at finite temperature can be better understood by recalling magnetic order arising in two-dimensional Ising model, which is a prototypical example of self-correcting \emph{classical} memory. The model has two degenerate ground states which are separated by $O(L)$ energy barrier where $L$ is the linear length of the lattice. At sufficiently low temperature, namely below the transition temperature $T_{c}$, the system is magnetically ordered, and a single bit of classical information may be securely stored. The memory time below $T_{c}$ can be estimated roughly by
\begin{align}
\tau \sim \exp\Big(\frac{E_{B} }{K_{B}T}\Big) \label{eq:Arr}
\end{align}
where $E_{B}$ is an energy barrier between two degenerate ground states and $T$ is the temperature. This predicts exponentially diverging memory time $\tau\sim \exp(L)$ at low temperatures, and thus the Ising model is a self-correcting memory. This empirical formula in Eq.~(\ref{eq:Arr}) is often called the Arrhenius law and is widely used to predict time scales of stochastic processes such as chemical reaction time. 

Interestingly, in the Ising model, the correspondence between magnetic order and self-correction is known to be exact:
\ba{
\mbox{magnetic order ($T<T_{c}$)}\quad  \Rightarrow \quad \tau \sim \exp(L^{D-1})\\
\mbox{no order ($T > T_{c}$)} \quad \Rightarrow \quad \tau \sim \log(L) \quad \quad 
}
where $D>1$ is the spatial dimension. Indeed, the above relations have been proven mathematically for $D=2$~\cite{Martinelli} and confirmed for $D=\infty$ in the mean field model. Significant progresses have been made toward proof of the correspondence for $D>2$~\cite{Martinelli}. Exponentially diverging memory time below $T_{c}$ can be easily verified via numerical simulations. A logarithmic memory time above $T_{c}$ can be proven from a well-known theorem on the mixing time of the Gibbs ensemble with exponentially decaying correlation functions~\cite{Martinelli}.

A naturally arising question concerns whether the presence of magnetic order generally implies self-correcting behavior with exponentially diverging memory time. In this paper, we argue that magnetic order does not necessarily imply self-correcting behavior by analyzing memory time of a certain well-known model of condensed matter physics (see~\cite{Tomita02, Borisenko11, Ortiz12} and references therein). In particular, we argue that the $q$-state clock model is magnetically ordered below the magnetic transition temperature $T_{c}$, but may possess polynomially diverging memory time despite the fact that energy barrier between degenerate ground states is $O(L)$. This weak violation of the Arrhenius law results from the fact that energy barrier is effectively suppressed to $\log(L)$ above the percolation temperature $T_{p}$ but below $T_{c}$ for $q\geq 5$. 

\emph{Clock model --}
The $q$-state clock model is a generalization of the Ising model to $q$-state spins. Spin values $s_{j}=0,\cdots,q-1$ may be associated with angular variables $\theta_{j}=\frac{2\pi}{q}s_{j}$. The Hamiltonian of the clock model is 
\begin{align}
H = -\sum_{[i,j]\in n.n.} \cos(\theta_{i}-\theta_{j})
\end{align}
where $i,j$ represent locations of spins on an $L\times L$ two-dimensional lattice with periodic boundary conditions. Interaction terms are proportional to the inner product of magnetization vectors, and are ferromagnetic. The system has $q$ degenerate ground states which are separated by $O(L)$ energy barrier. The total magnetization $m$ and the total angular value $\theta$ is given by $m e^{i\theta} = \frac{1}{L^{2}} \sum_{j} m_{j}$ where $m_{j}\equiv e^{i 2\pi s_{j}/q}$. 

\begin{figure}[htb!]
\centering
\includegraphics[width=1.00\linewidth]{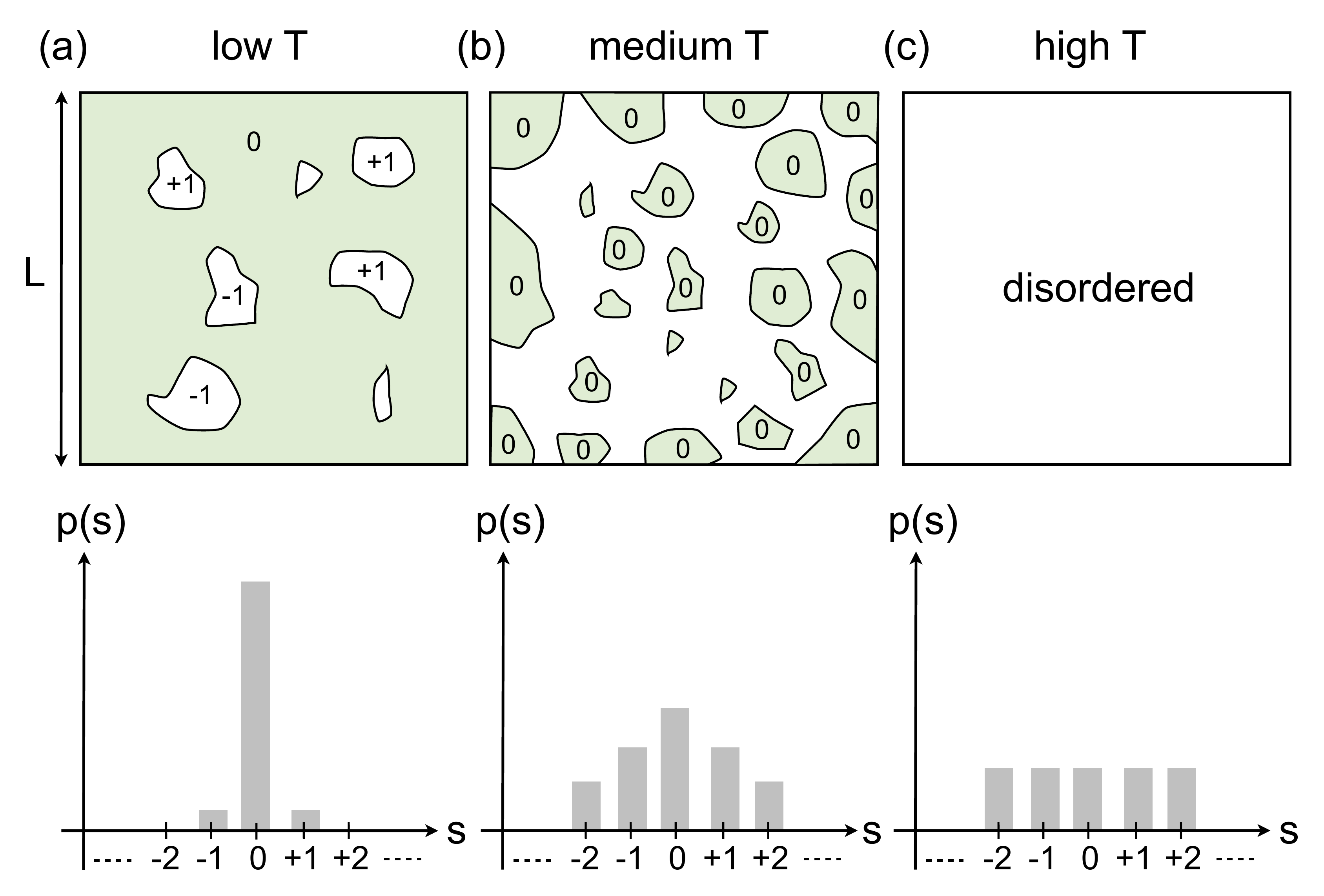}
\caption{Static properties of the clock model. (a) Low temperature phase. (b) Intermediate temperature phase. (c) High temperature phase.
} 
\label{fig_phase}
\end{figure}

In the low temperature regime, most of the spins are aligned in one particular direction. One may assume that most of the spins have spin value $s=0$. As depicted in Fig.~\ref{fig_phase}(a), one observes tiny excitation droplets which consist mostly of spins in $s=\pm 1$ since excitations of spins in $s=\pm 2$ cost larger energies and appear mostly inside islands of spins in $s=\pm1$. The probability distribution of spin species is sketched in Fig.~\ref{fig_phase}(a) which is centered around $s=0$. Therefore, one observes a finite magnetization $m>0$. 

If one increases the temperature, excitation droplets become larger and will eventually percolate the lattice (Fig.~\ref{fig_phase}(b)) at the percolation temperature $T_{p}$ where the islands of spins in $s=0$ are surrounded by excitations of spins in $s\not=0$. If this were the Ising model with $q=2$, the onset of percolation would mean the loss of total magnetization where the population of spins in $s=0$ becomes only half the numbers of spins. Yet, if $q$ is large, one still observes finite magnetization as the probability distribution is still centered around $s=0$ (Fig.~\ref{fig_phase}(b)). If we further increase the temperature, the system becomes magnetically disordered above the transition temperature $T_{c}$ and the probability distribution becomes uniform (Fig.~\ref{fig_phase}(c)). One can estimate the percolation temperature $T_{p}$ via the Peierls argument, which yields $T_{p}\approx \frac{4}{q^{2}}$, and the separation between $T_{c}$ and $T_{p}$ occurs for $q\geq 5$ (see~\cite{Ortiz12} for instance). 
 
\emph{Memory time --}
In the low temperature regime, the system loses the memory of its initial state if the excitation droplets grow as large as the system size. Since this process costs $O(L)$ energy, memory time is expected to be exponentially diverging $\sim \exp(L)$. Yet, in the intermediate temperature regime, $T_{p}<T<T_{c}$, in order for the system to lose the memory, one needs to reduce the number of islands of spins in $s=0$ which are already surrounded by excitations. Since the islands of spins in $s=0$ are mostly independent, the energy barrier is dominated by the energy required to eliminate the largest island of spins in $s=0$. According to the theory of percolation, at the percolation threshold, the sizes of percolated islands are at most logarithmic in the system size due to the scale invariance at criticality. thus, the energy barrier is effectively dropped to $\sim \log(L)$, and one obtains polynomially diverging memory time $\tau \sim \mbox{poly}(L)$ despite the fact that the system is magnetically ordered.

\begin{figure}[htb!]
\centering
\includegraphics[width=1.0\linewidth]{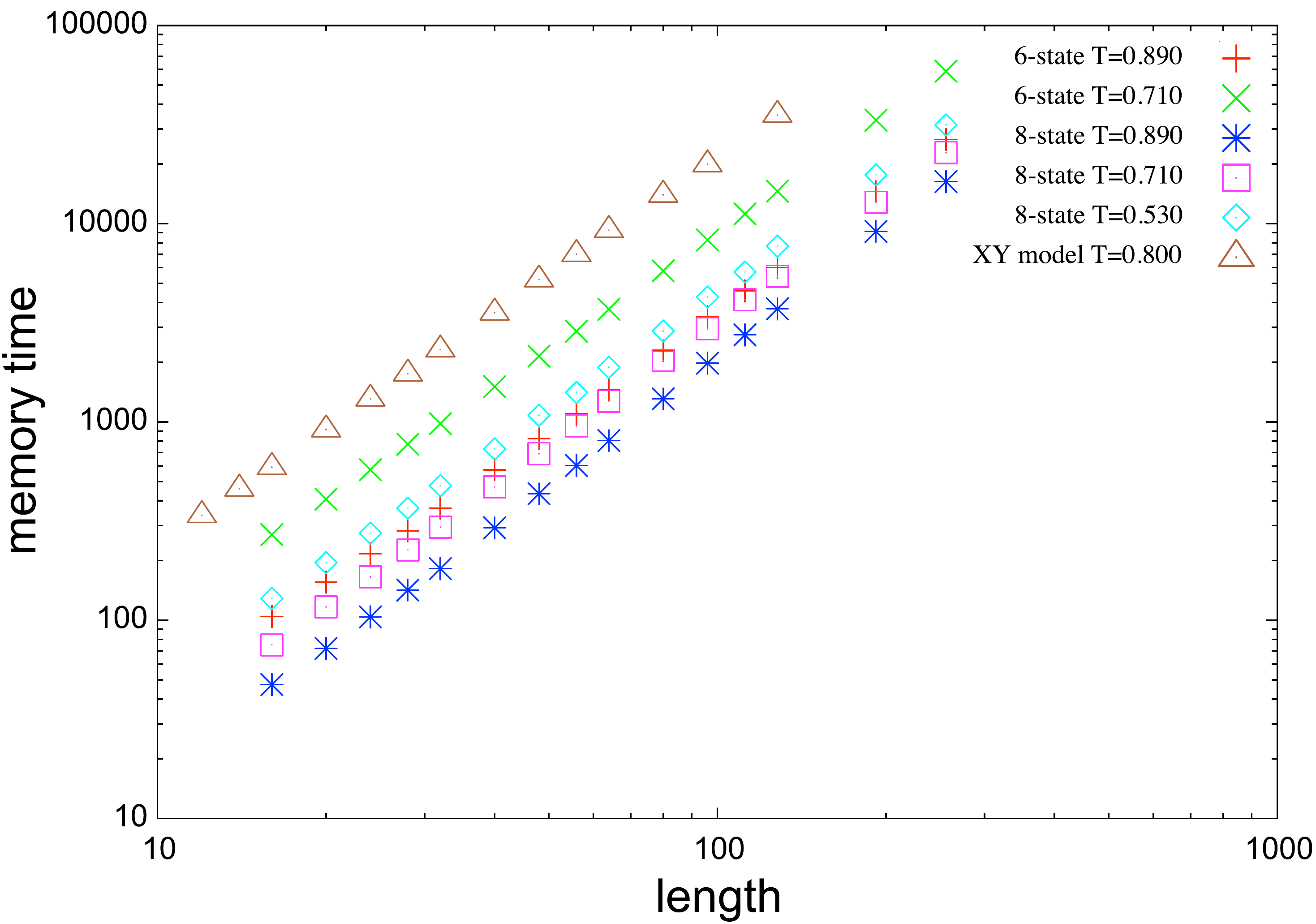}
\caption{Memory time of the clock model and the XY model versus length $L$ in a log-log plot. All the temperatures are in the intermediate regime.
} 
\label{fig_plot}
\end{figure}

We verify these theoretical predictions by numerical simulations. We first prepare a fully polarized state (say, all spins in $s=0$), couple it to the thermal bath and let it undergo thermal evolution, which is approximated by the Monte-Carlo simulation which implements the Metropolis algorithm. We compute the time for $s\not=0$ spins to become more dominant than spins in $s=0$. We take the average over $10,000$ realizations for $L\leq 128$ and $500$ realizations for $L=192,256$. The results for $6$-state and $8$-state clock models are plotted in a logarithmic scale in Fig.~\ref{fig_plot}(a). The plots can be fitted by straight lines, implying that the memory time is indeed polynomially diverging, with $\tau \propto L^{z}$.

One notices that slopes of the plots are similar, regardless of $q$ and $T$. This implies that the exponent $z$ is a universal quantity in the clock model. We find that $z \approx 2$, so the memory time is proportional to the volume. Recall that it takes $O(L^{2})$ time for a random walk to travel a distance of $O(L)$. As such, the fact $z\approx 2$ implies that the loss of memory time is a diffusive process instead of ballistic process. The memory time exponent $z$ seems to be related to the dynamical exponent $z'$ which governs the growth of correlation length under thermal quenches, $\xi \sim t^{1/z'}$, as often discussed in the context of the Kibble-Zurek mechanism. The dynamical exponent $z'$ of the clock model has been estimated by other methods, such as thermal evolution of correlation function, which also predicts $z' \approx 2$~\cite{Czerner96, Ozeki03, Corberi06}. 

\begin{figure}[htb!]
\centering
\includegraphics[width=0.95\linewidth]{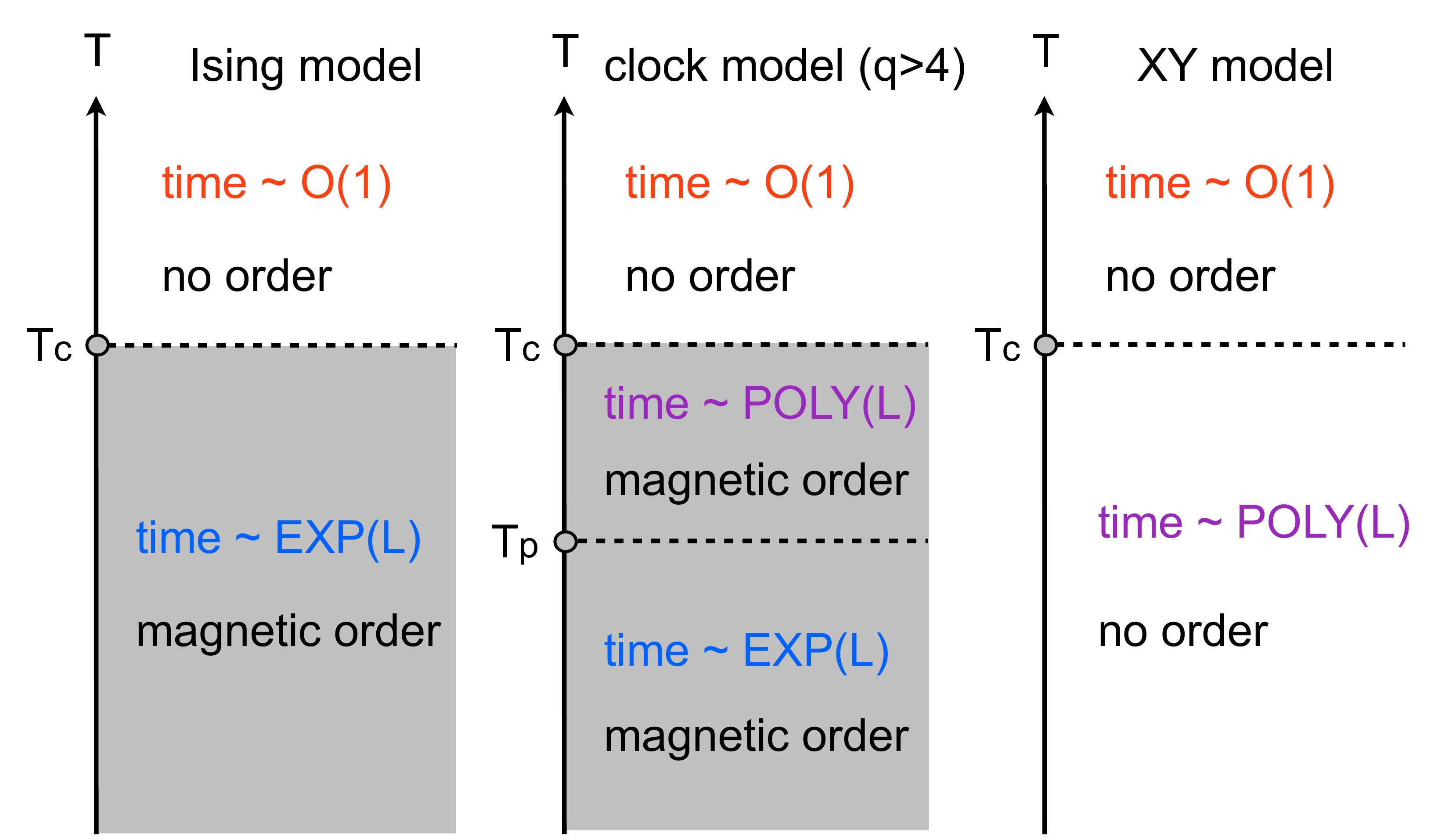}
\caption{Time scales of the clock model. Shaded regions represent phases with magnetic order. The system possesses magnetic order and polynomially diverging (experimentally relevant) time scale only for intermediate values of $q$.
} 
\label{fig_time}
\end{figure}

At $q\rightarrow \infty$ limit, the clock model is identical to two-dimensional XY model which undergoes a Berezinskii-Kosterlitz-Thouless transition at some finite temperature $T_{c}$ with no spontaneous symmetry-breaking. In the low temperature, $T<T_{c}$, the thermal dynamics is dominated by vortices whose energy grow $\sim \log(L)$. Despite the fact that total magnetization $m$ vanishes at any finite temperature, its memory time still diverges polynomially, as seen in Fig.~\ref{fig_time}.  

Below, we list estimates of exponents $z$ for various $q$ and $T$ in the intermediate temperature regime. For $q=2,3$, one has $z \approx 2.1$ while for $q=5$, one has $z \approx 1.9$. For $q>5$, the estimates of $z$ decreases as $T$ decreases. Note that for a fixed system size $L$ and $q$, the memory time must be a decreasing function of temperature $T$. This implies that an estimate of $z$ for high temperature must converge to an estimate of $z$ for low temperature at the thermodynamic limit ($L\rightarrow \infty$). Estimates at temperatures near $T_{p}$ all give $z \approx 2.0$ for $q\geq 6$. 

\begin{table}[h]
\centering
\begin{tabular}{c|c|c|c}
 state $q$ & temperature & exponent $z$  \\
\hline
$2$ & $T_{c}$ & 2.12(1)               \\
$3$ & $T_{c}$ & 2.11(1)                \\
$5$ & $0.950$ & 1.92(1)                 \\
$5$ & $0.910$ & 1.92(2)               \\
$6$ & $0.890$ & 2.11 (1)       \\
$6$ & $0.710$ & 2.00  (1)        \\
$8$ & $0.890$ &  2.13   (2)        \\
$8$ & $0.710$ & 2.07  (1)           \\
$8$ & $0.530$ & 2.04   (1)       \\
$8$ & $0.430$ & 1.98   (2)       \\
$12$ & $0.890$ & 2.61(5)           \\
$12$ & $0.550$ & 2.30(8)           \\
$12$ & $0.200$ &1.99   (3)     \\
$\infty$ & $0.800$ & 1.96  (1)         \\
\end{tabular}
\caption{Exponents of memory time for the clock model and the XY model.}
\end{table}

\emph{Time scales --}
We have seen that, even when the system is magnetically ordered below $T_{c}$, it may not work as a self-correcting memory due to the effective suppression of the energy barrier. Yet, one might think that polynomially diverging memory time is still useful for storing classical information. In order to answer this question, one needs to introduce actual time scales into discussion. 

The time scale of thermal dynamics depends on coupling strengths between the system and the thermal bath. In solid state physics, one Monte-Carlo step in a computer simulation typically corresponds to one pico-second ($10^{-12}$ second) in a real material~\cite{Belletti08}. For the Ising model (Fig.~\ref{fig_time}), at $T>T_{c}$, the memory time is extremely short, so the system loses its memory immediately, and reaches thermal equilibrium right after the contact with the thermal bath. At $T<T_{c}$, the memory time is $\sim \exp(L)$. Then, while one simulated Monte-Carlo step is an extremely fast process in a real material, the exponential growth is enormous, and the memory time easily exceeds the lifetime of the universe for $L\geq  100$. This implies that the system quickly reaches thermal equilibrium with spontaneously broken $\mathbb{Z}_{2}$ symmetry, and will remain in one of two magnetically ordered sectors forever. Unless $T = T_{c}$, thermal dynamics is too fast or too slow. Since the dynamical phase transition of time scales between two phases is sharp, in order to see dynamical phenomena at experimentally relevant time scales, one needs to tune the temperature to be $T=T_{c}$ precisely. 

In the clock model, there is a temperature regime with polynomially diverging memory time which may range from milliseconds to years depending on the system size $L$. This implies that, even though the system is magnetically ordered, its magnetization vector changes directions at experimentally relevant time scales, and the system does not reach the thermal equilibrium. This argument suggests an intriguing possibility of experimentally observing dynamical precession of magnetization vector in the percolated phase of the clock model. In order to verify this scenario, we perform a numerical simulation for the $6$-state clock model for $L=1024$ at an intermediate temperature $T=0.80$ ($T_{c}<T<T_{p}$). We observe that the total angular value $\theta$ indeed changes directions frequently, undergoing random walks while the total magnetization $m$ remains constant. From the scaling law of the memory time, $\tau \approx L^2$, one can estimate time scales of the precessions. For instance, for $L=10^6, 10^{8-9}$, the time scales are one second and one day respectively. These distinctions among three phases may allow us to view them as analogs of gas, liquid and solid, respectively, since only the liquid phase has dynamical phenomena at experimentally relevant time scale. There are a few materials, such as multiferroic hexagonal manganites, which may be effectively described by six-state clock model (see~\cite{Choi10} for instance).

\begin{figure}[htb!]
\centering
\includegraphics[width=0.95\linewidth]{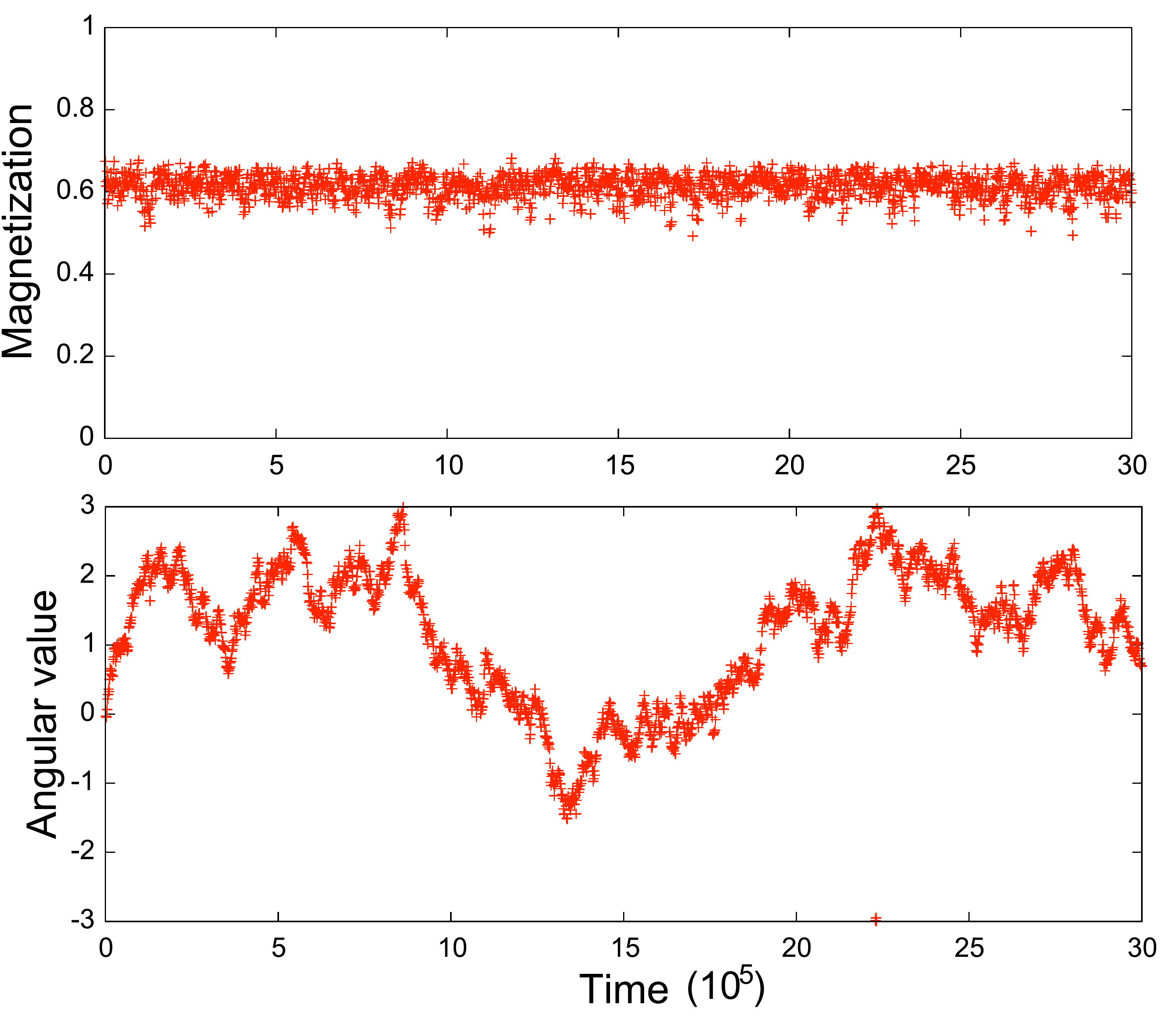}
\caption{Time evolutions of magnetization and polarization angle for $3\cdot 10^6$ Monte-Carlo steps. Polarization angles are converted into effective spin values $  \theta / 2\pi$ modulo $6$.
} 
\label{fig_precession}
\end{figure}

\emph{Discussion --}
In this paper, we have argued that finite $T_{c}$ does not necessarily imply self-correcting behaviors by microscopic analysis of the clock model. 
We have pointed out the possibility of experimentally observing precession of magnetization vectors in the clock model. At macroscopic level, only the polynomially diverging time scale seems to be experimentally relevant, and many of interesting dynamical phenomena indeed possess polynomially diverging time scales. Notable examples include the evaporation time of black holes and lifespan of animals versus their weight, known as the Kleiber's law~\cite{West97}. The common feature of all of these systems, including the clock model, is that they are out of equilibrium. Perhaps, information theoretical concepts, such as memory time and coding theory, may lead to a unified framework to study interesting non-equilibrium dynamics.

In summary, the following relations seem to hold:
\begin{equation}
\begin{split}
\mbox{magnetic order ($T<T_{c}$)}\quad  \Rightarrow \quad \tau \geq \mbox{poly}(L)\\
\mbox{no order ($T\geq T_{c}$)} \quad \Rightarrow \quad \tau \leq \mbox{poly}(L) \label{eq:relation}
\end{split}
\end{equation}
The $q$-state clock model ($q\geq 5$) and two-dimensional XY model ($q\rightarrow \infty$) correspond to the equalities in the above relations. A well-known result in mathematical physics community states that, if the system has exponentially decaying correlation function, it quickly reaches the thermal Gibbs state, and thus is not a good memory~\cite{Martinelli}. Presenting a rigorous proof of polynomial memory time in the clock model is also an important problem. It has been proven recently that memory time of two-dimensional Ising model is polynomially diverging at $T=T_{c}$~\cite{Lubetzky12}. Whether the above relations hold for more generic classical spin systems, such as gauge theories and classical fractal liquids~\cite{Beni11b}, is also an interesting open problem. It has been recently argued that higher-dimensional $\mathbb{Z}_{2}$ Toric code may work as a self-correcting quantum memory even above $T_{c}$ due to the hysteresis resulting from a first-order phase transition~\cite{Hastings14}. 

Generalization of Eq.~(\ref{eq:relation}) to the quantum setting is also an interesting problem. One can construct a quantum analog of the clock model by considering $\mathbb{Z}_{q}$ Toric code in higher dimensions. It may have an intermediate phase with polynomially diverging memory time while possessing nonzero topological entanglement entropy~\footnote{Non-zero topological entanglement entropy does not necessarily imply the presence of topological order at finite temperature since classical mixed states may also have finite topological entropy at finite temperature. For a proper definition, one needs to introduce a circuit complexity of preparing a Gibbs state as in~\cite{Hastings11}.}. Also, two-dimensional $\mathbb{Z}_{q}$ Toric code under active error-corrections may have polynomially diverging memory time below the error threshold~\cite{Dennis02}. Perhaps, the definition of error threshold needs some reconsideration as its current definition in quantum information community does not differentiate between exponential and polynomial memory time. Developing an efficient decoder for intermediate phases is also an interesting open question. 

\emph{Acknowledgment --}
I thank Alex Kubica, Fernando Pastawski and Kristan Temme for comments on the manuscript. I am supported by the David and Ellen Lee Postdoctoral fellowship. I acknowledge funding provided by the Institute for Quantum Information and Matter, an NSF Physics Frontiers Center with support of the Gordon and Betty Moore Foundation (Grants No. PHY-0803371 and PHY-1125565). This work used the Extreme Science and Engineering Discovery Environment (XSEDE), which is supported by National Science Foundation grant number OCI-1053575.


\end{document}